\documentclass[%
aip,
 amsmath,amssymb,
 reprint,%
]{revtex4-1}
\usepackage{graphicx}
\usepackage{dcolumn}
\usepackage{bm}
\usepackage{mathtools}
\usepackage{siunitx}
\usepackage{gensymb}
\usepackage{mathrsfs}
\usepackage{subfig}

\usepackage{amsmath}
\usepackage[utf8]{inputenc}
\usepackage[T1]{fontenc}
\usepackage{mathptmx}
\usepackage{etoolbox}
\usepackage{romannum}
\usepackage{xcolor}
\usepackage{natbib}
\usepackage{balance}
\usepackage{caption}
\makeatletter
\def\@email#1#2{%
 \endgroup
 \patchcmd{\titleblock@produce}
  {\frontmatter@RRAPformat}
  {\frontmatter@RRAPformat{\produce@RRAP{*#1\href{mailto:#2}{#2}}}\frontmatter@RRAPformat}
  {}{}
}%
\makeatother
\begin{document}

\preprint{AIP/123-QED}

\title{Vanadium doped $\beta$-Ga$_2$O$_3$ single crystals: Growth, Optical and Terahertz characterization}
\newcommand*{\affaddr}[1][]{#1} 
\newcommand*{\affmark}[1][*]{\textsuperscript{#1}}

\author{%
Maneesha Narayanan\affmark[1], Ajinkya Punjal\affmark[1,2], Emroj Hossain\affmark[1], Shraddha Choudhary\affmark[1], Ruta Kulkarni\affmark[1], S S Prabhu\affmark[1] Arumugam Thamizhavel\affmark[1] and Arnab Bhattacharya\affmark[1]\\
\affaddr{\affmark[1]\footnotesize \textit{Department of Condensed Matter Physics and Materials Science, Tata Institute of Fundamental Research (TIFR), Colaba, Mumbai 400005, India}}\\
\affaddr{\affmark[2]\footnotesize \textit{Department of Mechanical Engineering, Vishwakarma Institute of Information Technology (VIIT), Pune
411048, India}}%
}

\date{\today}

\begin{abstract}
We report the growth of electrically-resistive vanadium-doped $\beta$-Ga$_2$O$_3$ single crystals via the optical floating zone technique. By carefully controlling the growth parameters V-doped crystals with very high electrical resistivity compared to the usual $n$- type V-doped $\beta$-Ga$_2$O$_3$ (${n_e\approx10^{18}}$/cm$^{-3}$) can be synthesized. The optical properties of such high resistive V-doped $\beta$-Ga$_2$O$_3$ are significantly different compared to the undoped and $n$-doped crystals. We study the polarization-dependent Raman spectra, polarization-dependent transmission, temperature-dependent photoluminescence in the optical wavelength range and the THz transmission properties in the 0.2 - 2.6 THz range. The V-doped insulating Ga$_2$O$_3$ crystals show strong birefringence with refractive index contrast $\Delta$n of 0.3$\pm$ 0.02 at 1 THz, suggesting it to be an ideal material for optical applications in the THz region.

\end{abstract}

\maketitle

\section{\label{sec:level1}Introduction\protect\\}
Doping a semiconductor with suitable elements changes its physical, electrical, and optical properties\cite{7041161}${^,}$\cite{CHEN2021100369}. Undoped $\beta$-gallium oxide (Ga$_2$O$_3$) is a transparent, colourless and wide-bandgap semiconductor\cite{stepanov2016gallium} with a smallest direct gap of 4.87 eV.

It has a high breakdown ﬁeld\cite{adom} of 8 MV/cm  and high thermal stability\cite{Pearton1,doi:10.1063/1.4927742}. These properties make Ga$_2$O$_3$ a promising material for an extensive range of applications, such as high-power electronic devices \cite{doi:10.1063/5.0056557,doi:10.1063/5.0073005,doi:10.1063/1.3674287} and solar-blind ultraviolet (UV) photodetectors \cite{Tadjer_2018}${^,}$\cite{JI2006415}. While a theoretical study of transition metal (TM) doping in $\beta$-Ga$_2$O$_3$ has been reported\cite{gao2021effect}, it has been difficult to introduce dopants into a wide bandgap semiconductor like Ga$_2$O$_3$. The feasibility of doping Ga$_2$O$_3$ with vanadium in RF magnetron sputtered thin films has been studied\cite{HUANG201970}, the film quality obtained was rather mediocre. The synthesis of bulk single crystals of vanadium-doped gallium oxide has not been reported so far.

In this work we present the synthesis of high-quality Vanadium-doped insulating Ga$_2$O$_3$ by the optical floating zone (OFZ) technique. We report detailed, structural characterization, optical and Raman spectroscopic measurements, and preliminary terahertz time-doman spectroscopy (THz-TDS) data. 
\vspace{-5pt}
\vspace{-5pt}
\section{\label{sec:level2}Crystal Growth\protect\\}
The optical ﬂoating zone technique (OFZ) is a crucible-free method that allows the growth of very high-quality single crystals\cite{KOOHPAYEH2008121}. We have previously demonstrated high-quality OFZ grown undoped $\beta$-Ga$_2$O$_3$ single crystals\cite{Hossain_2019}. Details of our four-mirror OFZ growth system, and the process for the synthesis of (100) oriented $\beta$-Ga$_2$O$_3$ single crystal is presented in the Supporting Information, section 1 (SI\textendash\Romannum{1}). 

In this paper we call the unintentionally-doped $\beta$-Ga$_2$O$_3$  single crystal as undoped crystal. Undoped $\beta$-Ga$_2$O$_3$  single crystals are grown under standard conditions (an argon/oxygen ambient (90\% Ar and 10\% $O_2$) at atmospheric pressure, SI\textendash\Romannum{1}) show n-type conductivity with a free charge carrier concentration in the mid-$10^{17}cm^{-3}$ range determined from Hall mobility measurements. The free carrier concentration in the crystal can be controlled by growth conditions, post growth treatment or intentional doping. 

For the growth of vanadium doped crystals, 5N pure commercially available $\beta$-Ga$_2$O$_3$ powder was mixed with 1 mol\% of  99.2\% pure vanadium (\Romannum{5}) oxide powder. Since the OFZ growth zone is at a very high temperature ($\approx$ 1800\degree C), the thermal behaviour of dopant materials must be taken into account. This is even more critical for transition-metal elements like vanadium that exhibit variable oxidation states. V$_2$O$_5$ melts at 690\degree C and boils at 1750\degree C. Differential scanning calorimetry (DSC) measurements showed evidence for a phase change of V$_2$O$_5$ at 1180\degree C (Details are given in the SI\textendash\Romannum{2}). The typical process for doping in OFZ crystal growth involves sintering the precursor oxide powders at high temperature. To obtain insulating crystals, the precursor powder mixture was first sintered at 1250\degree C for 12hrs and then ground to a fine powder. A feed rod was made from this powder mixture which was again annealed at 1400\degree C for 24hrs. The crystal was grown over a [100] oriented seed crystal, in an argon/oxygen ambient (90\% Ar and 10\% $O_2$) at 1 atm pressure, at ﬁxed feed-seed rotation (10 rpm) and pulling rates (10 mm/hr). The sintering and annealing steps are critical in determining the resistivity of the final crystal. 
\vspace{-5pt}
\section{\label{sec:level3}Results and Discussions}

$\beta$-Ga$_2$O$_3$ has a base-centered monoclinic structure with C2/m space group symmetry\cite{geller}. Detailed crystal structure of $\beta$-Ga$_2$O$_3$ is presented the in SI\textendash\Romannum{3}. Comparing the ionic radii of vanadium with the ionic radii of the $Ga^{3+}$ ions, we believe they occupy the larger octahedral ($Ga_{\Romannum{2}} O_6$) sites in the unit cell.

\begin{figure}[h]
    \includegraphics[width =8.5cm]{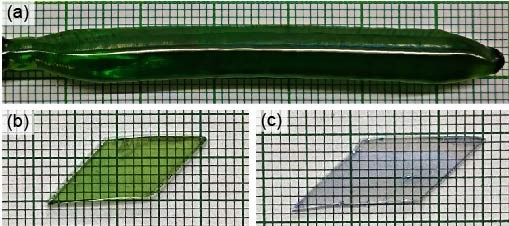}
    \caption{(a) Vanadium doped $\beta$-Ga$_2$O$_3$ single crystal grown in OFZ furnace. (b)-(c)  V-doped and undoped gallium oxide single crystals cleaved along (100) plane.}
    \label{fig:crystal}
\end{figure}

The V-doped $\beta$-Ga$_2$O$_3$ crystal is transparent  and green in color as shown in fig:\ref{fig:crystal}. 
The grown crystals easily cleave along [100] direction, giving "slices" as shown in (fig:\ref{fig:crystal}c). The structural, optical and terahertz properties of these crystals were measured on such slices of 0.3-0.8 mm in thickness, and compared with that of undoped crystal grown under same growth conditions. 

\begin{figure}
    \includegraphics[width=8.5cm]{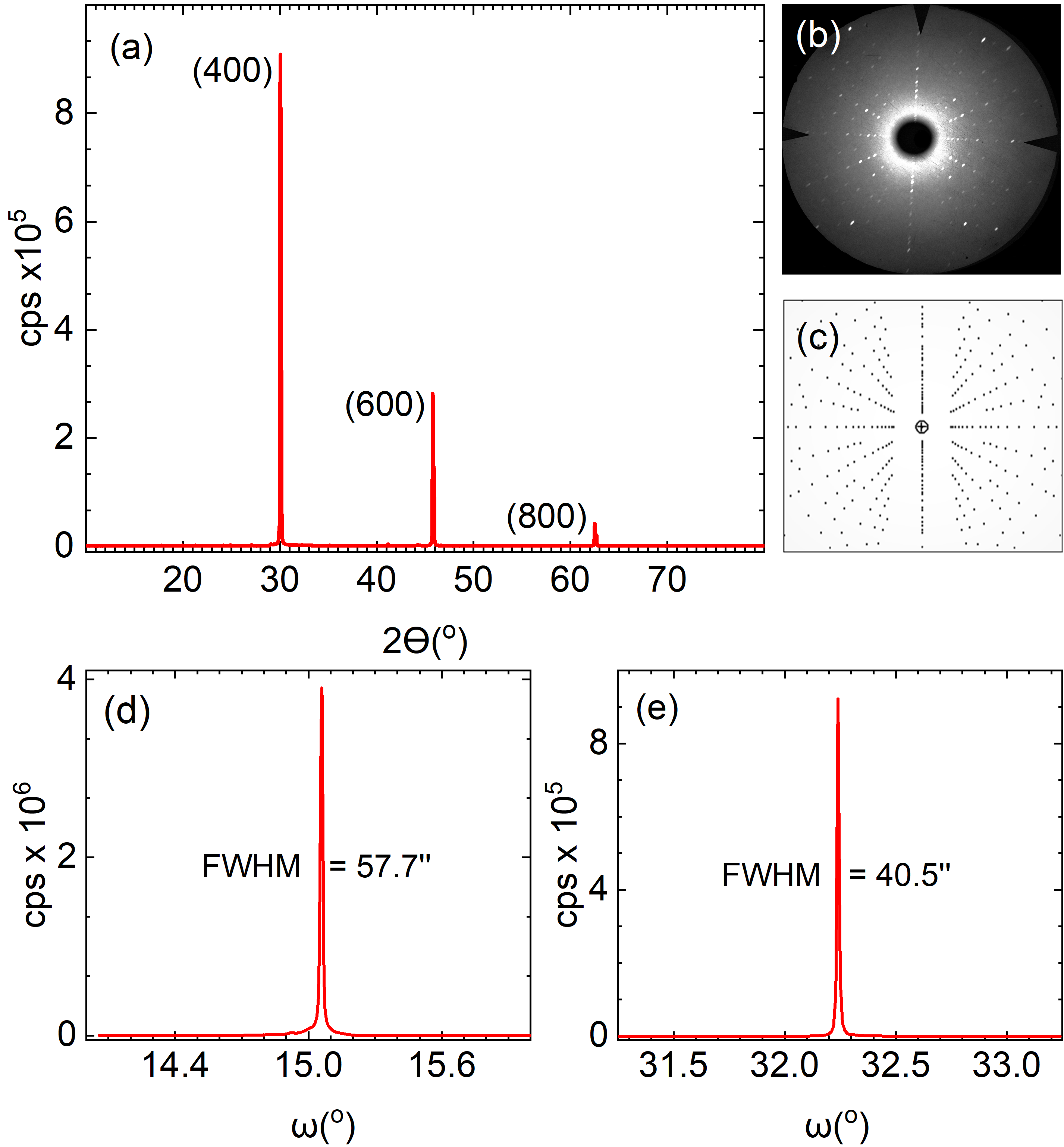}
    \caption{Structural characterization of the V-doped  $\beta$-Ga$_2$O$_3$ single crystals. (a) 2$\Theta$ scan for the cleaved sample along (100) plane. (b) experimental and (c) simulated Laue diffraction patterns. (d)-(e) Rocking curves along symmetric [100] axis and asymmetric [7 1 -2] axis respectively.}
    \label{fig:xrd}
\end{figure}

To check the crystallinity, various x-ray diffraction scans are used. The Laue and $\Theta$-$2\Theta$ scans are shown in fig:\ref{fig:xrd}. Both the doped and undoped crystals are of immensely high quality, with narrow rocking curves. Rocking curves for the V-doped crystal are shown in Fig. \ref{fig:xrd}(a-b). Symmetric and asymmetric scans were performed for (100) and (71-2) planes have FWHM values of 57.7'' and 40.5'', respectively.

To estimate the electrical conductivity Hall-effect measurements in the van-der-Pauw geometry were carried out at room temperature, using Ti and Au metal contacts annealed at 450\degree C for 1 minute. While undoped $\beta$-Ga$_2$O$_3$ samples were $n$-type and had a background charge carrier concentration $\approx$ $10^{17}$cm$^{-3}$, the V-doped samples were highly resistive (resistance > 200 M$\ohm$). Quantitative analysis using energy dispersive x-ray spectroscopy in a SEM showed the presence of 0.13-0.15 atomic\% vanadium in the crystal (Data is shown in SI\textendash\Romannum{4}). 
\begin{figure*}
    \includegraphics[width=17cm]{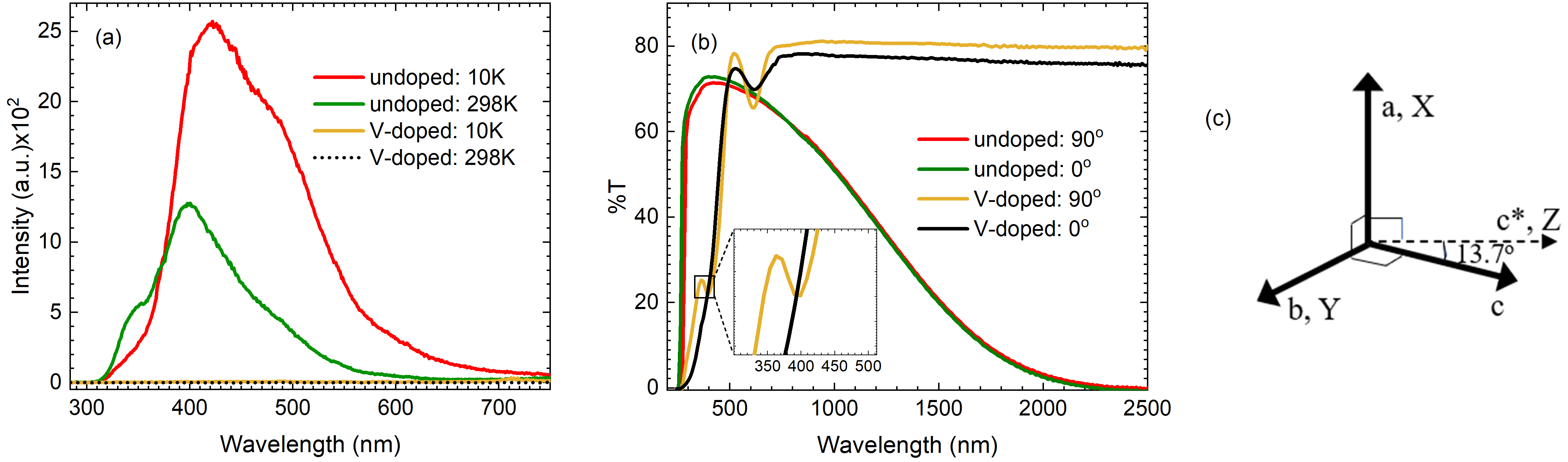}
    \caption{Optical characterization of undoped and V-doped  $\beta$-Ga$_2$O$_3$ single crystals. (a) Photoluminescence at 10 K and 298 K from V-doped and undoped $\beta$-Ga$_2$O$_3$ . The undoped crystal shows the typical broad defect luminescence band. However no observable luminescence is seen from the V-doped crystal at both temperatures. (b) Optical transmittance of V-doped and undoped crystals in two different orientations, $E$ $\parallel$ $b$ \& $E\perp$ $b$. (c) Crystal axis orientations ($a$, $b$ and $c$) with respect to incident laser/light source and signal paths (X, Y, and Z). For transmission measurement, the electric field is along Y for $E$ $\parallel$ $b$ and along Z for $E$ $\perp$ $b$ geometries.}
    \label{fig:images}
\end{figure*}

The photoluminescence (PL) spectrum of $\beta$-Ga$_2$O$_3$ does not exhibit near band-edge emission (NBE), but arises mostly from defect states in the material\cite{photoluminescence} in the spectral range 2-4 eV. Temperature dependent PL measurements for both undoped and doped samples were measured in the range 250 nm\textendash 750 nm with a 266 nm laser excitation source. There was no PL signal at higher wavelength for both the crystals. Fig:\ref{fig:images}(a) shows photoluminescence from V-doped and undoped $\beta$-Ga$_2$O$_3$  at low temperature (10 K) and room temperature. The undoped crystal shows the typical broad emission spectrum associated with defect states. The luminescence is surprisingly quenched in the V-doped single crystal. Transition metal dopants are known to give rise to deep level states within the gap that serve as non-radiative recombination pathways that can kill the luminescence typically seen in undoped materials. The vanadium atoms act as compensating acceptor dopants, which also explains the semi-insulating nature of the material.

The structure of $\beta$-Ga$_2$O$_3$ with Ga-O chains aligned along the $b$ axis\cite{HARWIG1978255, BINET19981241} gives rise to an intrinsic anisotropy in the electrical and optical properties for measurements made parallel and perpendicular to this axis. Room temperature optical transmittance of the undoped and doped crystals (cleaved slices of $\approx$ 0.3 mm thickness) at two different polarizations \textendash ~$E$ $\parallel$ $b$ (0\degree) and $E$ $\perp$ $b$ (90\degree) \textendash  ~were measured using a V-670 Jasco Spectrometer. The transmittance spectra in the wavelength range 200 nm to 2500 nm with a source change at 340 nm and grating change at 850 nm, and are shown in fig:\ref{fig:images}(b). With its wide bandgap of 4.87 eV, undoped gallium oxide transmits $\approx$70\% of incident light in the visible region and transmission gradually decreases towards longer wavelength side, due to free carrier absorption. However, depending on oxygen vacancies and deep levels in the bandgap created by the dopant atoms, the absorption in the crystal can vary.
Transmittance for doped crystal decreased largely in the UV region (300 nm\textendash450 nm). Also, there are two broad absorption features centered at 390 nm and 615 nm arising from inter bandgap absorption. For bandgap analysis we need a much thin sample which is difficult to obtain due to the cleaving nature of the crystal in (001) plane. The transmission spectra for the undoped and V-doped crystals are  polarization dependent with the doped crystal having a larger difference ($\approx$4-5\%  between the $E$ $\parallel$ $b$ \& $E\perp$ $b$ directions). The V-doped crystal transmits 77-80\% of incident signal in the green and NIR spectral regions and 65-70\% in the red region of visible spectrum. The horizontal nature of transmittance for the doped sample suggests the absence of free carriers, which also supports the insulating nature of the crystal. For both samples, we observe that the transmittance traces for the $E$ $\parallel$ $b$ \& $E\perp$ $b$ directions cross each other, indicating that the crystals show optical birefringence. A comparison of the spectra also indicates that the samples exhibit a higher band edge energy for the $E\parallel$ $b$ polarization direction.

\begin{figure*}
    \includegraphics[width=17cm]{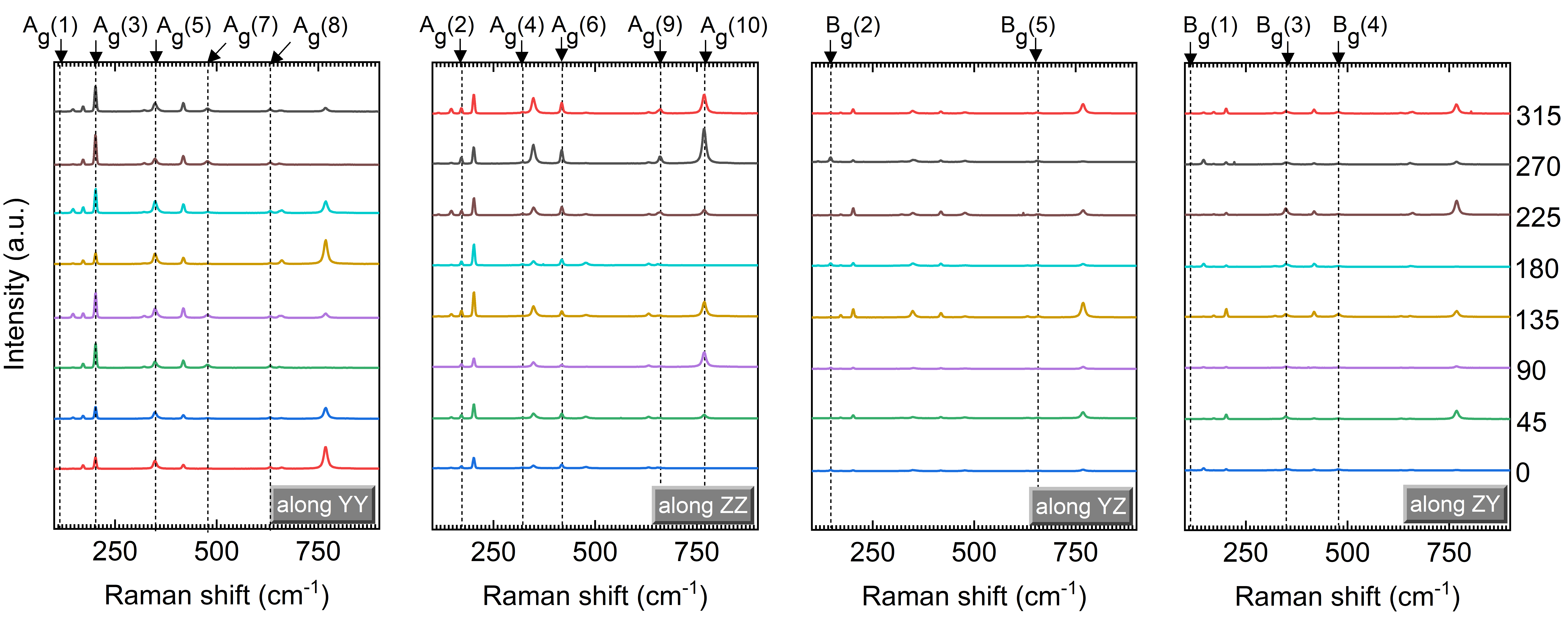}
    \caption{Polarized raman spectra under planar rotations for the V-doped sample. The different polarization orientations for the doped sample are given. 180\degree symmetry is observed.}
    \label{fig:raman}
\end{figure*}

There are 30 vibrational modes \cite{ONUMA2014330} in $\beta$-Ga$_2$O$_3$. Among these 30 modes, 3 are acoustic and 27 are optical. The irreducible representations for acoustic modes for $\beta$-Ga$_2$O$_3$ is given by\cite{doi:10.1063/5.0059070}
\begin{equation}
    \centering
    \Gamma_{aco} = A_u + 2 B_u 
    \label{eqn:modes}
\end{equation}
and for optical modes
\begin{equation}
    \centering
    \Gamma_{opt} = 10 A_g + 5 B_g + 4 A_u + 8 B_u
    \label{eqn:modes}
\end{equation}
While the $A_g$ and $B_g$ modes are Raman active, $A_u$ and  $B_u$ modes are infrared active. Raman measurements were performed using a WITec $\alpha$ - 300 confocal Raman system with a 1800 rulings/mm gratings and 532 nm laser excitation source. Raman shifts for both the undoped and doped samples were almost identical (SI\textendash\Romannum{5}). Polarization dependent Raman measurements were also done by rotating the (100) oriented sample in the azimuthal plane, to obtain an angular dependence of the scattering intensity (fig:\ref{fig:raman}). The X-axis is taken along [100] direction and is perpendicular to the cleaved surface. The Y-direction is taken along [010]-direction. The angle between the crystallographic a and c axis is 103.7$\degree$ resulting in an offset of 13.7$\degree$ between crystallographic c and Z axis (fig:\ref{fig:images}(c)). The measurement is done on the sample in four different polarization directions - YY, ZZ, YZ, and ZY, (fig:\ref{fig:raman}) where the former character represents the direction of the incident electric field vector and the latter represents the direction of the electric field vector in the scattered light. The excitation of vibrational modes depends on the polarization selection rule\cite{polarizationselection}. YY and ZZ conﬁgurations allow $A_g$ modes, and YZ and YZ conﬁgurations allow $B_g$ modes to dominate.

While the angular dependence is expected to be identical for angles 180 $\degree$ apart due to the symmetry of the crystal, there are small differences in the relative intensity of the Raman peaks, which may arise from local strain induced by the dopant atoms. The Raman peaks in Ga$_2$O$_3$ can be divided in to three parts\cite{ONUMA2014330}. Libration and translation of chains Ga(\Romannum{1})O(\Romannum{1}) chains( $A_g(1)- A_g(3),B_g(1)$ and $B_g(2)$) ,deformation of Ga(\Romannum{1})O(\Romannum{2}/\Romannum{3}) chains ( $A_g(4)-A_g(6),B_g(3)$ and $B_g(4)$) , stretching and bending of Ga(\Romannum{1})O(\Romannum{1}) ($A_g(7)-A_g(10)$ and $B_g(5)$). The angular dependence of the intensities under planar rotations is similar for modes belonging to the same category. For modes in different categories, e.g. $A_g(3)$ and $A_g(10)$, there is a 90\degree shift in the angular positions where the intensity is a maximum as the crystal is rotated.
  
\begin{figure*}
    \includegraphics[width=17cm]{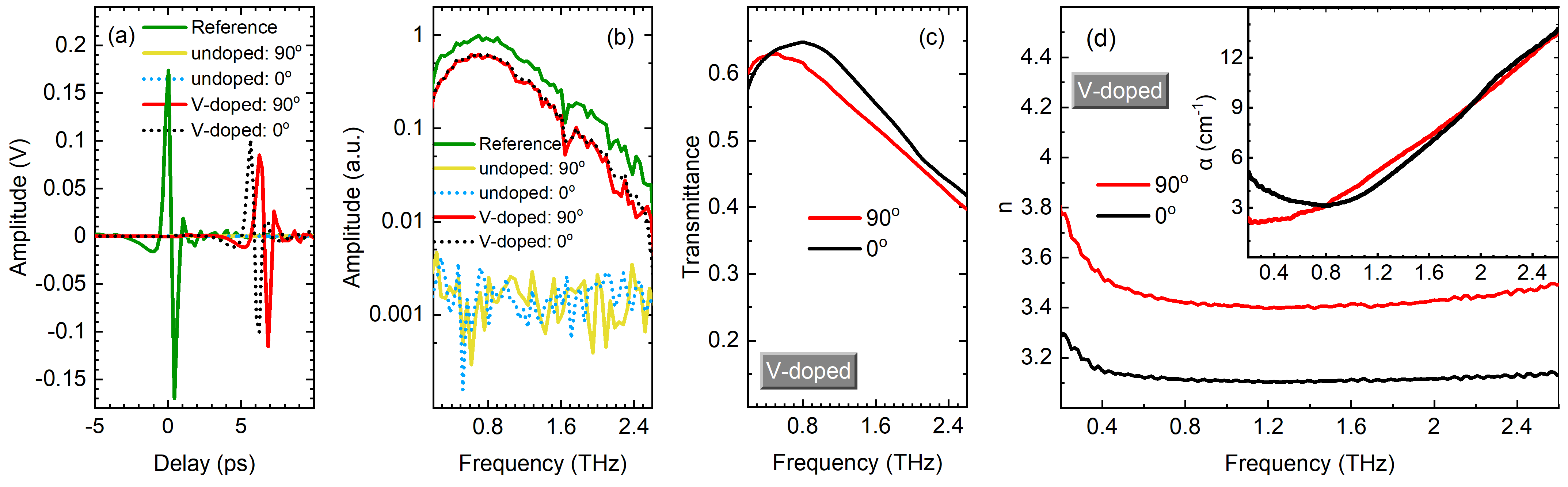}
   \caption{(a) The transmitted THz transients along ($0^o$) and ($90^o$) for both the undoped and V-doped Ga$_2$O$_3$ crystals. (b) Corresponding Fourier transformed spectra. (c) THz transmittance of V-doped crystal along two different orientations. (d) Refractive Index ($n_S$) in the range 0.2-2.6 THz is given with Absorption coefficient [$\alpha (cm^{-1})$] plot in the inset, for two different orientations.}
    \label{fig:THz}
\end{figure*}

Terahertz time-domain spectroscopy (THz-TDS) of the samples were performed using a lab-built setup, details of which are given in the SI\textendash\Romannum{6}. The measurements were performed along $E$ $\parallel$ $b$ \& $E\perp$ $b$ directions for studying the anisotropic properties. The obtained THz transients of the undoped and V-doped Ga$_2$O$_3$ crystals are shown in Fig:\ref{fig:THz}(a). While the V-doped crystal transmits a reasonable amount of the THz signal through it, transmission through the undoped crystal is negligible. This further supports the insulating nature of V-doped crystal as well as the conducting nature of undoped crystal. A shift in transmitted signal is observed for $E$ $\parallel$ $b$ \& $E\perp$ $b$ directions of the V-doped crystal, which indicates a difference in optical path lengths through the crystal, suggesting different refractive indices along 0\degree and 90\degree rotation of the crystal, the hallmark property of a birefringent material. The measured temporal THz wave forms were Fourier transformed and the resultant spectra for both the undoped and V-doped crystals are shown in Fig:\ref{fig:THz}(b). The complex transmittance coefficients were extracted from this data, and the real part is plotted in Fig:\ref{fig:THz}(c). As seen, the overall transmittance is more than 50\% from 0.4 to 1.8 THz. The real part of the refractive index ($n$) is calculated from the phase ($\phi(\omega)$) of transmittance using the following relation\cite{doi:10.1063/1.5047659, Duvillaret:99}.
\begin{equation}
    n(\omega) \ = \ 1-\frac{c}{\omega d} \phi(\omega)\\
    \label{eqn:Thz1}
\end{equation}
where d is the thickness of the sample, c the speed of light in vacuum, and $\omega$ the frequency. 
Fig:\ref{fig:THz}(d) shows the V-doped crystal has two different refractive indices along 0\degree and 90\degree. The refractive index for 0\degree and 90\degree at 1 THz is $3.10\pm0.01$ and $3.40\pm0.01$, respectively, giving a refractive index contrast of $\Delta n = |0.3\pm0.02|$. To our knowledge, vanadium doping shows the highest THz birefringence in  Ga$_2$O$_3$ crystals, doped or undoped, reported to date~\cite{doi:10.1063/5.0031531,agulto2021anisotropic,blumenschein2020dielectric}. 
The imaginary part of the refractive index (extinction coefficient ($k$) is calculated from the amplitude of the transmittance $\lvert$T($\omega$)$\rvert$,

\begin{equation}
   k(\omega) \ = \ \frac{c}{\omega d}\left[{ln\frac{4n(\omega)}{[n(\omega)+1]^2} - ln\lvert T(\omega) \rvert}\right] 
    \label{eqn:Thz2}
\end{equation}

which further gives information about the absorption coefficient ($\alpha$) of the crystal,
\begin{equation}
   \alpha(\omega)= 2 K k(\omega)
    \label{eqn:Thz3}
\end{equation}
Where $K$ is the wave number. The absorption coefficient for both orientations (inset of Fig:\ref{fig:THz}(d)) is nearly 4 $cm^{-1}$ at 1 THz, which is similar to that of silica glass\cite{naftaly2007terahertz} ($\approx$ 2.6 $cm^{-1}$) at 1 THz.

\section{Conclusion}
Vanadium doping significantly alters the electrical and optical properties of $\beta$-Ga$_2$O$_3$. Vanadium atoms give rise to deep level states within the gap leading to non-radiative recombination pathways that suppress the photoluminescence typically seen in undoped materials. The V-doped Ga$_2$O$_3$ crystals are strongly birefringent in the THz region from 0.2 to 2.6 THz. The overall contrast in the refractive index is $\Delta n=|0.3\pm0.02|$ at 1 THz. This, coupled with their electrically resistive property makes V-doped Ga$_2$O$_3$ crystals are ideal candidate for exploring potential terahertz applications such as single wavelength quarter wave plate, half wave plate, segmented wave plates and faraday rotators.

\section{Author Declarations}
The authors have no conflicts to disclose

\section{Author Contributions}
The crystals were grown by MN and RK, with structural and optical characterizations performed by MN, under the overall guidance of AT and AB. EH provided useful and critical inputs for crystal growth and characterization. AP and SC conducted terahertz spectroscopy under the supervision of SSP. The manuscript was written by MN and SC, and reviewed by all authors.

\section{Acknowledgements}
The authors thank Azizur Rahman, Mahesh Gokhale and Amit Shah for help in Raman, XRD and electrical measurements, Prof. Sandip Ghosh, Vishwas Jindal and Dibya Sankar Das for discussions and help with PL measurements, Bhagyashree A Chalke, Rudeer D Bapat and Jayesh B Parmar for EDX analyses and Kushal Darje and the TIFR central workshop for making accessories for the growth system. This research was funded by Department of Atomic Energy (DAE) grant RTI4003.

\section{DATA AVAILABILITY STATEMENT}
The data of this study are available from the corresponding author upon reasonable request.

\bibliographystyle{aapmrev4-1} 
\bibliography{aipsamp} 

\begin{thebibliography}{28}%
\makeatletter
\providecommand \@ifxundefined [1]{%
 \@ifx{#1\undefined}
}%
\providecommand \@ifnum [1]{%
 \ifnum #1\expandafter \@firstoftwo
 \else \expandafter \@secondoftwo
 \fi
}%
\providecommand \@ifx [1]{%
 \ifx #1\expandafter \@firstoftwo
 \else \expandafter \@secondoftwo
 \fi
}%
\providecommand \natexlab [1]{#1}%
\providecommand \enquote  [1]{``#1''}%
\providecommand \bibnamefont  [1]{#1}%
\providecommand \bibfnamefont [1]{#1}%
\providecommand \citenamefont [1]{#1}%
\providecommand \href@noop [0]{\@secondoftwo}%
\providecommand \href [0]{\begingroup \@sanitize@url \@href}%
\providecommand \@href[1]{\@@startlink{#1}\@@href}%
\providecommand \@@href[1]{\endgroup#1\@@endlink}%
\providecommand \@sanitize@url [0]{\catcode `\\12\catcode `\$12\catcode
  `\&12\catcode `\#12\catcode `\^12\catcode `\_12\catcode `\%12\relax}%
\providecommand \@@startlink[1]{}%
\providecommand \@@endlink[0]{}%
\providecommand \url  [0]{\begingroup\@sanitize@url \@url }%
\providecommand \@url [1]{\endgroup\@href {#1}{\urlprefix }}%
\providecommand \urlprefix  [0]{URL }%
\providecommand \Eprint [0]{\href }%
\providecommand \doibase [0]{http://dx.doi.org/}%
\providecommand \selectlanguage [0]{\@gobble}%
\providecommand \bibinfo  [0]{\@secondoftwo}%
\providecommand \bibfield  [0]{\@secondoftwo}%
\providecommand \translation [1]{[#1]}%
\providecommand \BibitemOpen [0]{}%
\providecommand \bibitemStop [0]{}%
\providecommand \bibitemNoStop [0]{.\EOS\space}%
\providecommand \EOS [0]{\spacefactor3000\relax}%
\providecommand \BibitemShut  [1]{\csname bibitem#1\endcsname}%
\let\auto@bib@innerbib\@empty
\bibitem [{\citenamefont {Chang}\ \emph {et~al.}(2015)\citenamefont {Chang},
  \citenamefont {Chang}, \citenamefont {Chiu}, \citenamefont {Wei},
  \citenamefont {Juan},\ and\ \citenamefont {Weng}}]{7041161}%
  \BibitemOpen
  \bibfield  {author} {\bibinfo {author} {\bibfnamefont {T.-H.}\ \bibnamefont
  {Chang}}, \bibinfo {author} {\bibfnamefont {S.-J.}\ \bibnamefont {Chang}},
  \bibinfo {author} {\bibfnamefont {C.~J.}\ \bibnamefont {Chiu}}, \bibinfo
  {author} {\bibfnamefont {C.-Y.}\ \bibnamefont {Wei}}, \bibinfo {author}
  {\bibfnamefont {Y.-M.}\ \bibnamefont {Juan}}, \ and\ \bibinfo {author}
  {\bibfnamefont {W.-Y.}\ \bibnamefont {Weng}},\ }\href@noop {} {\bibfield
  {journal} {\bibinfo  {journal} {IEEE Photonics Technology Letters}\ }\textbf
  {\bibinfo {volume} {27}},\ \bibinfo {pages} {915} (\bibinfo {year}
  {2015})}\BibitemShut {NoStop}%
\bibitem [{\citenamefont {Chen}\ \emph {et~al.}(2021)\citenamefont {Chen},
  \citenamefont {Lu}, \citenamefont {Yang}, \citenamefont {Li}, \citenamefont
  {Li}, \citenamefont {Chen}, \citenamefont {Xu}, \citenamefont {Zang},\ and\
  \citenamefont {Shan}}]{CHEN2021100369}%
  \BibitemOpen
  \bibfield  {author} {\bibinfo {author} {\bibfnamefont {Y.}~\bibnamefont
  {Chen}}, \bibinfo {author} {\bibfnamefont {Y.}~\bibnamefont {Lu}}, \bibinfo
  {author} {\bibfnamefont {X.}~\bibnamefont {Yang}}, \bibinfo {author}
  {\bibfnamefont {S.}~\bibnamefont {Li}}, \bibinfo {author} {\bibfnamefont
  {K.}~\bibnamefont {Li}}, \bibinfo {author} {\bibfnamefont {X.}~\bibnamefont
  {Chen}}, \bibinfo {author} {\bibfnamefont {Z.}~\bibnamefont {Xu}}, \bibinfo
  {author} {\bibfnamefont {J.}~\bibnamefont {Zang}}, \ and\ \bibinfo {author}
  {\bibfnamefont {C.}~\bibnamefont {Shan}},\ }\href@noop {} {\bibfield
  {journal} {\bibinfo  {journal} {Materials Today Physics}\ }\textbf {\bibinfo
  {volume} {18}},\ \bibinfo {pages} {100369} (\bibinfo {year}
  {2021})}\BibitemShut {NoStop}%
\bibitem [{\citenamefont {Stepanov}\ \emph {et~al.}(2016)\citenamefont
  {Stepanov}, \citenamefont {Nikolaev}, \citenamefont {Bougrov},\ and\
  \citenamefont {Romanov}}]{stepanov2016gallium}%
  \BibitemOpen
  \bibfield  {author} {\bibinfo {author} {\bibfnamefont {S.}~\bibnamefont
  {Stepanov}}, \bibinfo {author} {\bibfnamefont {V.}~\bibnamefont {Nikolaev}},
  \bibinfo {author} {\bibfnamefont {V.}~\bibnamefont {Bougrov}}, \ and\
  \bibinfo {author} {\bibfnamefont {A.}~\bibnamefont {Romanov}},\ }\href@noop
  {} {\bibfield  {journal} {\bibinfo  {journal} {Rev. Adv. Mater. Sci}\
  }\textbf {\bibinfo {volume} {44}},\ \bibinfo {pages} {63} (\bibinfo {year}
  {2016})}\BibitemShut {NoStop}%
\bibitem [{\citenamefont {Deng}\ \emph {et~al.}(2020)\citenamefont {Deng},
  \citenamefont {Leedle}, \citenamefont {Miao}, \citenamefont {Black},
  \citenamefont {Urbanek}, \citenamefont {McNeur}, \citenamefont {Kozák},
  \citenamefont {Ceballos}, \citenamefont {Hommelhoff}, \citenamefont
  {Solgaard}, \citenamefont {Byer},\ and\ \citenamefont {Harris}}]{adom}%
  \BibitemOpen
  \bibfield  {author} {\bibinfo {author} {\bibfnamefont {H.}~\bibnamefont
  {Deng}}, \bibinfo {author} {\bibfnamefont {K.~J.}\ \bibnamefont {Leedle}},
  \bibinfo {author} {\bibfnamefont {Y.}~\bibnamefont {Miao}}, \bibinfo {author}
  {\bibfnamefont {D.~S.}\ \bibnamefont {Black}}, \bibinfo {author}
  {\bibfnamefont {K.~E.}\ \bibnamefont {Urbanek}}, \bibinfo {author}
  {\bibfnamefont {J.}~\bibnamefont {McNeur}}, \bibinfo {author} {\bibfnamefont
  {M.}~\bibnamefont {Kozák}}, \bibinfo {author} {\bibfnamefont
  {A.}~\bibnamefont {Ceballos}}, \bibinfo {author} {\bibfnamefont
  {P.}~\bibnamefont {Hommelhoff}}, \bibinfo {author} {\bibfnamefont
  {O.}~\bibnamefont {Solgaard}}, \bibinfo {author} {\bibfnamefont {R.~L.}\
  \bibnamefont {Byer}}, \ and\ \bibinfo {author} {\bibfnamefont {J.~S.}\
  \bibnamefont {Harris}},\ }\href@noop {} {\bibfield  {journal} {\bibinfo
  {journal} {Advanced Optical Materials}\ }\textbf {\bibinfo {volume} {8}},\
  \bibinfo {pages} {1901522} (\bibinfo {year} {2020})}\BibitemShut {NoStop}%
\bibitem [{\citenamefont {Pearton}(2018)}]{Pearton1}%
  \BibitemOpen
  \bibfield  {author} {\bibinfo {author} {\bibfnamefont {p.~H.~c.}\
  \bibnamefont {Pearton}, \bibfnamefont {Jiancheng~Yang}},\ }\href@noop {}
  {\bibfield  {journal} {\bibinfo  {journal} {Applied Physics Reviews}\
  }\textbf {\bibinfo {volume} {5}},\ \bibinfo {pages} {011301} (\bibinfo {year}
  {2018})}\BibitemShut {NoStop}%
\bibitem [{\citenamefont {Santia}, \citenamefont {Tandon},\ and\ \citenamefont
  {Albrecht}(2015)}]{doi:10.1063/1.4927742}%
  \BibitemOpen
  \bibfield  {author} {\bibinfo {author} {\bibfnamefont {M.~D.}\ \bibnamefont
  {Santia}}, \bibinfo {author} {\bibfnamefont {N.}~\bibnamefont {Tandon}}, \
  and\ \bibinfo {author} {\bibfnamefont {J.~D.}\ \bibnamefont {Albrecht}},\
  }\href@noop {} {\bibfield  {journal} {\bibinfo  {journal} {Applied Physics
  Letters}\ }\textbf {\bibinfo {volume} {107}},\ \bibinfo {pages} {041907}
  (\bibinfo {year} {2015})}\BibitemShut {NoStop}%
\bibitem [{\citenamefont {Chatterjee}\ \emph {et~al.}(2021)\citenamefont
  {Chatterjee}, \citenamefont {Li}, \citenamefont {Nomoto}, \citenamefont
  {Xing},\ and\ \citenamefont {Choi}}]{doi:10.1063/5.0056557}%
  \BibitemOpen
  \bibfield  {author} {\bibinfo {author} {\bibfnamefont {B.}~\bibnamefont
  {Chatterjee}}, \bibinfo {author} {\bibfnamefont {W.}~\bibnamefont {Li}},
  \bibinfo {author} {\bibfnamefont {K.}~\bibnamefont {Nomoto}}, \bibinfo
  {author} {\bibfnamefont {H.~G.}\ \bibnamefont {Xing}}, \ and\ \bibinfo
  {author} {\bibfnamefont {S.}~\bibnamefont {Choi}},\ }\href@noop {} {\bibfield
   {journal} {\bibinfo  {journal} {Applied Physics Letters}\ }\textbf {\bibinfo
  {volume} {119}},\ \bibinfo {pages} {103502} (\bibinfo {year}
  {2021})}\BibitemShut {NoStop}%
\bibitem [{\citenamefont {Zheng}, \citenamefont {Zhao},\ and\ \citenamefont
  {Feng}(2022)}]{doi:10.1063/5.0073005}%
  \BibitemOpen
  \bibfield  {author} {\bibinfo {author} {\bibfnamefont {X.-Q.}\ \bibnamefont
  {Zheng}}, \bibinfo {author} {\bibfnamefont {H.}~\bibnamefont {Zhao}}, \ and\
  \bibinfo {author} {\bibfnamefont {P.~X.-L.}\ \bibnamefont {Feng}},\
  }\href@noop {} {\bibfield  {journal} {\bibinfo  {journal} {Applied Physics
  Letters}\ }\textbf {\bibinfo {volume} {120}},\ \bibinfo {pages} {040502}
  (\bibinfo {year} {2022})}\BibitemShut {NoStop}%
\bibitem [{\citenamefont {Higashiwaki}\ \emph {et~al.}(2012)\citenamefont
  {Higashiwaki}, \citenamefont {Sasaki}, \citenamefont {Kuramata},
  \citenamefont {Masui},\ and\ \citenamefont
  {Yamakoshi}}]{doi:10.1063/1.3674287}%
  \BibitemOpen
  \bibfield  {author} {\bibinfo {author} {\bibfnamefont {M.}~\bibnamefont
  {Higashiwaki}}, \bibinfo {author} {\bibfnamefont {K.}~\bibnamefont {Sasaki}},
  \bibinfo {author} {\bibfnamefont {A.}~\bibnamefont {Kuramata}}, \bibinfo
  {author} {\bibfnamefont {T.}~\bibnamefont {Masui}}, \ and\ \bibinfo {author}
  {\bibfnamefont {S.}~\bibnamefont {Yamakoshi}},\ }\href@noop {} {\bibfield
  {journal} {\bibinfo  {journal} {Applied Physics Letters}\ }\textbf {\bibinfo
  {volume} {100}},\ \bibinfo {pages} {013504} (\bibinfo {year}
  {2012})}\BibitemShut {NoStop}%
\bibitem [{\citenamefont {Tadjer}(2018)}]{Tadjer_2018}%
  \BibitemOpen
  \bibfield  {author} {\bibinfo {author} {\bibfnamefont {M.~J.}\ \bibnamefont
  {Tadjer}},\ }\href@noop {} {\bibfield  {journal} {\bibinfo  {journal} {The
  Electrochemical Society Interface}\ }\textbf {\bibinfo {volume} {27}},\
  \bibinfo {pages} {49} (\bibinfo {year} {2018})}\BibitemShut {NoStop}%
\bibitem [{\citenamefont {Ji}\ \emph {et~al.}(2006)\citenamefont {Ji},
  \citenamefont {Du}, \citenamefont {Fan},\ and\ \citenamefont
  {Wang}}]{JI2006415}%
  \BibitemOpen
  \bibfield  {author} {\bibinfo {author} {\bibfnamefont {Z.}~\bibnamefont
  {Ji}}, \bibinfo {author} {\bibfnamefont {J.}~\bibnamefont {Du}}, \bibinfo
  {author} {\bibfnamefont {J.}~\bibnamefont {Fan}}, \ and\ \bibinfo {author}
  {\bibfnamefont {W.}~\bibnamefont {Wang}},\ }\href@noop {} {\bibfield
  {journal} {\bibinfo  {journal} {Optical Materials}\ }\textbf {\bibinfo
  {volume} {28}},\ \bibinfo {pages} {415} (\bibinfo {year} {2006})}\BibitemShut
  {NoStop}%
\bibitem [{\citenamefont {Gao}\ \emph {et~al.}(2021)\citenamefont {Gao},
  \citenamefont {Li}, \citenamefont {Dai}, \citenamefont {Wang},\ and\
  \citenamefont {Suo}}]{gao2021effect}%
  \BibitemOpen
  \bibfield  {author} {\bibinfo {author} {\bibfnamefont {S.}~\bibnamefont
  {Gao}}, \bibinfo {author} {\bibfnamefont {W.}~\bibnamefont {Li}}, \bibinfo
  {author} {\bibfnamefont {J.}~\bibnamefont {Dai}}, \bibinfo {author}
  {\bibfnamefont {Q.}~\bibnamefont {Wang}}, \ and\ \bibinfo {author}
  {\bibfnamefont {Z.}~\bibnamefont {Suo}},\ }\href@noop {} {\bibfield
  {journal} {\bibinfo  {journal} {Materials Research Express}\ }\textbf
  {\bibinfo {volume} {8}},\ \bibinfo {pages} {025904} (\bibinfo {year}
  {2021})}\BibitemShut {NoStop}%
\bibitem [{\citenamefont {Huang}\ \emph {et~al.}(2019)\citenamefont {Huang},
  \citenamefont {Hu}, \citenamefont {Zou}, \citenamefont {Tang}, \citenamefont
  {Zhang}, \citenamefont {Ma}, \citenamefont {Li}, \citenamefont {Wang},\ and\
  \citenamefont {Lu}}]{HUANG201970}%
  \BibitemOpen
  \bibfield  {author} {\bibinfo {author} {\bibfnamefont {J.}~\bibnamefont
  {Huang}}, \bibinfo {author} {\bibfnamefont {Y.}~\bibnamefont {Hu}}, \bibinfo
  {author} {\bibfnamefont {T.}~\bibnamefont {Zou}}, \bibinfo {author}
  {\bibfnamefont {K.}~\bibnamefont {Tang}}, \bibinfo {author} {\bibfnamefont
  {Z.}~\bibnamefont {Zhang}}, \bibinfo {author} {\bibfnamefont
  {Y.}~\bibnamefont {Ma}}, \bibinfo {author} {\bibfnamefont {B.}~\bibnamefont
  {Li}}, \bibinfo {author} {\bibfnamefont {L.}~\bibnamefont {Wang}}, \ and\
  \bibinfo {author} {\bibfnamefont {Y.}~\bibnamefont {Lu}},\ }\href@noop {}
  {\bibfield  {journal} {\bibinfo  {journal} {Surface and Coatings Technology}\
  }\textbf {\bibinfo {volume} {366}},\ \bibinfo {pages} {70} (\bibinfo {year}
  {2019})}\BibitemShut {NoStop}%
\bibitem [{\citenamefont {Koohpayeh}, \citenamefont {Fort},\ and\ \citenamefont
  {Abell}(2008)}]{KOOHPAYEH2008121}%
  \BibitemOpen
  \bibfield  {author} {\bibinfo {author} {\bibfnamefont {S.}~\bibnamefont
  {Koohpayeh}}, \bibinfo {author} {\bibfnamefont {D.}~\bibnamefont {Fort}}, \
  and\ \bibinfo {author} {\bibfnamefont {J.}~\bibnamefont {Abell}},\
  }\href@noop {} {\bibfield  {journal} {\bibinfo  {journal} {Progress in
  Crystal Growth and Characterization of Materials}\ }\textbf {\bibinfo
  {volume} {54}},\ \bibinfo {pages} {121} (\bibinfo {year} {2008})}\BibitemShut
  {NoStop}%
\bibitem [{\citenamefont {Hossain}(2019)}]{Hossain_2019}%
  \BibitemOpen
  \bibfield  {author} {\bibinfo {author} {\bibfnamefont {E.}~\bibnamefont
  {Hossain}},\ }\href@noop {} {\bibfield  {journal} {\bibinfo  {journal} {The
  Electrochemical Society}\ }\textbf {\bibinfo {volume} {8}},\ \bibinfo {pages}
  {Q3144} (\bibinfo {year} {2019})}\BibitemShut {NoStop}%
\bibitem [{\citenamefont {Geller}(1960)}]{geller}%
  \BibitemOpen
  \bibfield  {author} {\bibinfo {author} {\bibfnamefont {S.}~\bibnamefont
  {Geller}},\ }\href@noop {} {\bibfield  {journal} {\bibinfo  {journal} {The
  journal of Chemical Physics}\ }\textbf {\bibinfo {volume} {33}},\ \bibinfo
  {pages} {676} (\bibinfo {year} {1960})}\BibitemShut {NoStop}%
\bibitem [{\citenamefont {Frodason}\ \emph {et~al.}(2020)\citenamefont
  {Frodason}, \citenamefont {Johansen}, \citenamefont {Vines},\ and\
  \citenamefont {Varley}}]{photoluminescence}%
  \BibitemOpen
  \bibfield  {author} {\bibinfo {author} {\bibfnamefont {Y.~K.}\ \bibnamefont
  {Frodason}}, \bibinfo {author} {\bibfnamefont {K.~M.}\ \bibnamefont
  {Johansen}}, \bibinfo {author} {\bibfnamefont {L.}~\bibnamefont {Vines}}, \
  and\ \bibinfo {author} {\bibfnamefont {J.~B.}\ \bibnamefont {Varley}},\
  }\href@noop {} {\bibfield  {journal} {\bibinfo  {journal} {Journal of Applied
  Physics}\ }\textbf {\bibinfo {volume} {127}},\ \bibinfo {pages} {075701}
  (\bibinfo {year} {2020})}\BibitemShut {NoStop}%
\bibitem [{\citenamefont {Harwig}\ and\ \citenamefont
  {Kellendonk}(1978)}]{HARWIG1978255}%
  \BibitemOpen
  \bibfield  {author} {\bibinfo {author} {\bibfnamefont {T.}~\bibnamefont
  {Harwig}}\ and\ \bibinfo {author} {\bibfnamefont {F.}~\bibnamefont
  {Kellendonk}},\ }\href@noop {} {\bibfield  {journal} {\bibinfo  {journal}
  {Journal of Solid State Chemistry}\ }\textbf {\bibinfo {volume} {24}},\
  \bibinfo {pages} {255} (\bibinfo {year} {1978})}\BibitemShut {NoStop}%
\bibitem [{\citenamefont {Binet}\ and\ \citenamefont
  {Gourier}(1998)}]{BINET19981241}%
  \BibitemOpen
  \bibfield  {author} {\bibinfo {author} {\bibfnamefont {L.}~\bibnamefont
  {Binet}}\ and\ \bibinfo {author} {\bibfnamefont {D.}~\bibnamefont
  {Gourier}},\ }\href@noop {} {\bibfield  {journal} {\bibinfo  {journal}
  {Journal of Physics and Chemistry of Solids}\ }\textbf {\bibinfo {volume}
  {59}},\ \bibinfo {pages} {1241} (\bibinfo {year} {1998})}\BibitemShut
  {NoStop}%
\bibitem [{\citenamefont {Onuma}\ \emph {et~al.}(2014)\citenamefont {Onuma},
  \citenamefont {Fujioka}, \citenamefont {Yamaguchi}, \citenamefont {Itoh},
  \citenamefont {Higashiwaki}, \citenamefont {Sasaki}, \citenamefont {Masui},\
  and\ \citenamefont {Honda}}]{ONUMA2014330}%
  \BibitemOpen
  \bibfield  {author} {\bibinfo {author} {\bibfnamefont {T.}~\bibnamefont
  {Onuma}}, \bibinfo {author} {\bibfnamefont {S.}~\bibnamefont {Fujioka}},
  \bibinfo {author} {\bibfnamefont {T.}~\bibnamefont {Yamaguchi}}, \bibinfo
  {author} {\bibfnamefont {Y.}~\bibnamefont {Itoh}}, \bibinfo {author}
  {\bibfnamefont {M.}~\bibnamefont {Higashiwaki}}, \bibinfo {author}
  {\bibfnamefont {K.}~\bibnamefont {Sasaki}}, \bibinfo {author} {\bibfnamefont
  {T.}~\bibnamefont {Masui}}, \ and\ \bibinfo {author} {\bibfnamefont
  {T.}~\bibnamefont {Honda}},\ }\href@noop {} {\bibfield  {journal} {\bibinfo
  {journal} {Journal of Crystal Growth}\ }\textbf {\bibinfo {volume} {401}},\
  \bibinfo {pages} {330} (\bibinfo {year} {2014})}\BibitemShut {NoStop}%
\bibitem [{\citenamefont {Onuma}\ \emph {et~al.}(2021)\citenamefont {Onuma},
  \citenamefont {Sasaki}, \citenamefont {Yamaguchi}, \citenamefont {Honda},
  \citenamefont {Kuramata}, \citenamefont {Yamakoshi},\ and\ \citenamefont
  {Higashiwaki}}]{doi:10.1063/5.0059070}%
  \BibitemOpen
  \bibfield  {author} {\bibinfo {author} {\bibfnamefont {T.}~\bibnamefont
  {Onuma}}, \bibinfo {author} {\bibfnamefont {K.}~\bibnamefont {Sasaki}},
  \bibinfo {author} {\bibfnamefont {T.}~\bibnamefont {Yamaguchi}}, \bibinfo
  {author} {\bibfnamefont {T.}~\bibnamefont {Honda}}, \bibinfo {author}
  {\bibfnamefont {A.}~\bibnamefont {Kuramata}}, \bibinfo {author}
  {\bibfnamefont {S.}~\bibnamefont {Yamakoshi}}, \ and\ \bibinfo {author}
  {\bibfnamefont {M.}~\bibnamefont {Higashiwaki}},\ }\href@noop {} {\bibfield
  {journal} {\bibinfo  {journal} {Applied Physics Letters}\ }\textbf {\bibinfo
  {volume} {118}},\ \bibinfo {pages} {252101} (\bibinfo {year}
  {2021})}\BibitemShut {NoStop}%
\bibitem [{\citenamefont {Kranert}\ \emph {et~al.}(2016)\citenamefont
  {Kranert}, \citenamefont {Christian}, \citenamefont {Chris}, \citenamefont
  {Rüdiger},\ and\ \citenamefont {Grundmann}}]{polarizationselection}%
  \BibitemOpen
  \bibfield  {author} {\bibinfo {author} {\bibnamefont {Kranert}}, \bibinfo
  {author} {\bibfnamefont {S.}~\bibnamefont {Christian}}, \bibinfo {author}
  {\bibfnamefont {S.-G.}\ \bibnamefont {Chris}}, \bibinfo {author}
  {\bibnamefont {Rüdiger}}, \ and\ \bibinfo {author} {\bibfnamefont
  {M.}~\bibnamefont {Grundmann}},\ }\href@noop {} {\bibfield  {journal}
  {\bibinfo  {journal} {Scientific Reports}\ }\textbf {\bibinfo {volume} {6}},\
  \bibinfo {pages} {35964} (\bibinfo {year} {2016})}\BibitemShut {NoStop}%
\bibitem [{\citenamefont {Neu}\ and\ \citenamefont
  {Schmuttenmaer}(2018)}]{doi:10.1063/1.5047659}%
  \BibitemOpen
  \bibfield  {author} {\bibinfo {author} {\bibfnamefont {J.}~\bibnamefont
  {Neu}}\ and\ \bibinfo {author} {\bibfnamefont {C.~A.}\ \bibnamefont
  {Schmuttenmaer}},\ }\href@noop {} {\bibfield  {journal} {\bibinfo  {journal}
  {Journal of Applied Physics}\ }\textbf {\bibinfo {volume} {124}},\ \bibinfo
  {pages} {231101} (\bibinfo {year} {2018})}\BibitemShut {NoStop}%
\bibitem [{\citenamefont {Duvillaret}, \citenamefont {Garet},\ and\
  \citenamefont {Coutaz}(1999)}]{Duvillaret:99}%
  \BibitemOpen
  \bibfield  {author} {\bibinfo {author} {\bibfnamefont {L.}~\bibnamefont
  {Duvillaret}}, \bibinfo {author} {\bibfnamefont {F.}~\bibnamefont {Garet}}, \
  and\ \bibinfo {author} {\bibfnamefont {J.-L.}\ \bibnamefont {Coutaz}},\
  }\href@noop {} {\bibfield  {journal} {\bibinfo  {journal} {Appl. Opt.}\
  }\textbf {\bibinfo {volume} {38}},\ \bibinfo {pages} {409} (\bibinfo {year}
  {1999})}\BibitemShut {NoStop}%
\bibitem [{\citenamefont {Agulto}\ \emph
  {et~al.}(2021{\natexlab{a}})\citenamefont {Agulto}, \citenamefont {Toya},
  \citenamefont {Phan}, \citenamefont {Mag-usara}, \citenamefont {Li},
  \citenamefont {Empizo}, \citenamefont {Iwamoto}, \citenamefont {Goto},
  \citenamefont {Murakami}, \citenamefont {Kumagai}, \citenamefont {Sarukura},
  \citenamefont {Yoshimura},\ and\ \citenamefont
  {Nakajima}}]{doi:10.1063/5.0031531}%
  \BibitemOpen
  \bibfield  {author} {\bibinfo {author} {\bibfnamefont {V.~C.}\ \bibnamefont
  {Agulto}}, \bibinfo {author} {\bibfnamefont {K.}~\bibnamefont {Toya}},
  \bibinfo {author} {\bibfnamefont {T.~N.~K.}\ \bibnamefont {Phan}}, \bibinfo
  {author} {\bibfnamefont {V.~K.}\ \bibnamefont {Mag-usara}}, \bibinfo {author}
  {\bibfnamefont {J.}~\bibnamefont {Li}}, \bibinfo {author} {\bibfnamefont
  {M.~J.~F.}\ \bibnamefont {Empizo}}, \bibinfo {author} {\bibfnamefont
  {T.}~\bibnamefont {Iwamoto}}, \bibinfo {author} {\bibfnamefont
  {K.}~\bibnamefont {Goto}}, \bibinfo {author} {\bibfnamefont {H.}~\bibnamefont
  {Murakami}}, \bibinfo {author} {\bibfnamefont {Y.}~\bibnamefont {Kumagai}},
  \bibinfo {author} {\bibfnamefont {N.}~\bibnamefont {Sarukura}}, \bibinfo
  {author} {\bibfnamefont {M.}~\bibnamefont {Yoshimura}}, \ and\ \bibinfo
  {author} {\bibfnamefont {M.}~\bibnamefont {Nakajima}},\ }\href@noop {}
  {\bibfield  {journal} {\bibinfo  {journal} {Applied Physics Letters}\
  }\textbf {\bibinfo {volume} {118}},\ \bibinfo {pages} {042101} (\bibinfo
  {year} {2021}{\natexlab{a}})}\BibitemShut {NoStop}%
\bibitem [{\citenamefont {Agulto}\ \emph
  {et~al.}(2021{\natexlab{b}})\citenamefont {Agulto}, \citenamefont {Toya},
  \citenamefont {Phan}, \citenamefont {Mag-usara}, \citenamefont {Li},
  \citenamefont {Empizo}, \citenamefont {Iwamoto}, \citenamefont {Goto},
  \citenamefont {Murakami}, \citenamefont {Kumagai} \emph
  {et~al.}}]{agulto2021anisotropic}%
  \BibitemOpen
  \bibfield  {author} {\bibinfo {author} {\bibfnamefont {V.~C.}\ \bibnamefont
  {Agulto}}, \bibinfo {author} {\bibfnamefont {K.}~\bibnamefont {Toya}},
  \bibinfo {author} {\bibfnamefont {T.~N.~K.}\ \bibnamefont {Phan}}, \bibinfo
  {author} {\bibfnamefont {V.~K.}\ \bibnamefont {Mag-usara}}, \bibinfo {author}
  {\bibfnamefont {J.}~\bibnamefont {Li}}, \bibinfo {author} {\bibfnamefont
  {M.~J.~F.}\ \bibnamefont {Empizo}}, \bibinfo {author} {\bibfnamefont
  {T.}~\bibnamefont {Iwamoto}}, \bibinfo {author} {\bibfnamefont
  {K.}~\bibnamefont {Goto}}, \bibinfo {author} {\bibfnamefont {H.}~\bibnamefont
  {Murakami}}, \bibinfo {author} {\bibfnamefont {Y.}~\bibnamefont {Kumagai}},
  \emph {et~al.},\ }\href@noop {} {\bibfield  {journal} {\bibinfo  {journal}
  {Applied Physics Letters}\ }\textbf {\bibinfo {volume} {118}},\ \bibinfo
  {pages} {042101} (\bibinfo {year} {2021}{\natexlab{b}})}\BibitemShut
  {NoStop}%
\bibitem [{\citenamefont {Blumenschein}\ \emph {et~al.}(2020)\citenamefont
  {Blumenschein}, \citenamefont {Kadlec}, \citenamefont {Romanyuk},
  \citenamefont {Paskova}, \citenamefont {Muth},\ and\ \citenamefont
  {Kadlec}}]{blumenschein2020dielectric}%
  \BibitemOpen
  \bibfield  {author} {\bibinfo {author} {\bibfnamefont {N.}~\bibnamefont
  {Blumenschein}}, \bibinfo {author} {\bibfnamefont {C.}~\bibnamefont
  {Kadlec}}, \bibinfo {author} {\bibfnamefont {O.}~\bibnamefont {Romanyuk}},
  \bibinfo {author} {\bibfnamefont {T.}~\bibnamefont {Paskova}}, \bibinfo
  {author} {\bibfnamefont {J.~F.}\ \bibnamefont {Muth}}, \ and\ \bibinfo
  {author} {\bibfnamefont {F.}~\bibnamefont {Kadlec}},\ }\href@noop {}
  {\bibfield  {journal} {\bibinfo  {journal} {Journal of Applied Physics}\
  }\textbf {\bibinfo {volume} {127}},\ \bibinfo {pages} {165702} (\bibinfo
  {year} {2020})}\BibitemShut {NoStop}%
\bibitem [{\citenamefont {Naftaly}\ and\ \citenamefont
  {Miles}(2007)}]{naftaly2007terahertz}%
  \BibitemOpen
  \bibfield  {author} {\bibinfo {author} {\bibfnamefont {M.}~\bibnamefont
  {Naftaly}}\ and\ \bibinfo {author} {\bibfnamefont {R.}~\bibnamefont
  {Miles}},\ }\href@noop {} {\bibfield  {journal} {\bibinfo  {journal} {Journal
  of Applied Physics}\ }\textbf {\bibinfo {volume} {102}},\ \bibinfo {pages}
  {043517} (\bibinfo {year} {2007})}\BibitemShut {NoStop}%
\end{thebibliography}%


%

\end{document}